# Universal Lattice Basis

Jonathan L Jerke and C. J. Tymczak

*Department of Physics, Texas Southern University, Houston, TX 77004, USA*

## Abstract

We report on the utility of using Shannon's Sampling theorem to solve Quantum Mechanical systems. We show that by extending the logic of Shannon's interpolation theorem we can define a Universal Lattice Basis, which has superior interpolating properties compared to traditional methods. This basis is orthonormal, semi-local, has a Euclidean norm, and a simple analytic expression for the derivatives. Additionally, we can define a bounded domain for which band-limited functions, such as Gaussians, show quadratic convergence in the representation error in respect to the sampling frequency. This theory also extends to the periodic domain and we illustrate the simple analytic forms of the periodic semi-local basis and derivatives. Additionally, we show that this periodic basis is equivalent to the space defined by the Fast Fourier Transform. This novel basis has great utility in solving quantum mechanical problems for which the wave functions are known to be naturally band-limited. Several numerical examples in single and multi-dimensions are given to show the convergence and equivalence of the periodic and bounded domains for compact states.

## I. Introduction

The work of information theory in the 1950's by Shannon et. el. [1, 2] discovered a means to reconstruct continuous data streams from a uniform sampling. Shannon's sampling theorem states that all Fourier components with wavelengths longer than *2d*, where *d* is the lattice spacing, are exact onto a lattice. However, this result from Shannon did not result in an expression for a derivative on the lattice. This paper will use this early result based on the *Sinc* function to express the lattice derivative with these limitations taken into account.

A lattice derivative equates the lattice image of the derivative with the lattice derivative of the lattice image. When short wavelengths are ignored (band-limited) there is a unique solution. Fast Fourier Transforms accomplish this via an $O(N)$ algorithm that efficiently transforms the coordinated space vector to a Fourier component spectra and back [3]. In what follows, we will present analytic expressions for the basis representation and lattice derivative for unbounded domains, bounded domains and periodic domains.

In references [4-6] Marston and then Kosloff solved the Schrödinger wave equations using Fast Fourier Transform techniques. Within their works they showed how to construct a derivative matrix using a summation over all the Fourier modes. These authors used a mixture of Fourier techniques for computation of the derivative and *Sinc* function to justify diagonalizing the potential in space. In references [7, 8] Yaroslavsky and then Blu considered the *Sinc* basis and attempted to make it more local with a modification based on a filter but then failed to derive an exact representation of the derivative. Campos et. al in reference [9] discussed a quadrature approximation to a derivative of a Hermite-Gaussian basis. To our knowledge there are no published works that use the *Sinc* basis given by Shannon interpolation formula as a basis space to solve the Schrödinger wave equations. We believe that this was caused by the difficulty in dealing with *Sinc's* functions semi non-locality.

This paper is organized as follows. In section II we will first describe the properties of our basis space, which we refer to as Lattice Universals, composed of the *Sinc* functions. In section III we will derive the extensions of this space into the periodic domain showing the connections to discrete Fourier space. In section IV we will discuses the computational and theoretical considerations needed to solve quantum mechanical problems, and in section V we will show the application of the basis to solving several quantum mechanical problems of general interest. Finally, in section VI we will give some concluding remarks and future directions.

## II. Lattice Universals

**Eikon:** We define an Eikon as a discrete representation of a continuous function. It is represented in coordinate space using only trigonometric functions such as the *Sinc* function. It is effectively the band-limited Fourier transform of the original continuous function.

**Cardinal Sine:** The *Sinc* function is the means to interpolate band-limited functions. The *Sinc* function that we consider is normalized to the lattice spacing, *d*, via,

$$Sinc(\vartheta) = \frac{1}{\sqrt{d}} \frac{Sin(\pi\vartheta)}{\pi\vartheta} \tag{1}$$

Where $\vartheta$ is the dimensionless parameter, which has a place of an index or sub index written in the form of *x/d*. The *Sinc* function is the kernel of the Universal Lattice Space. The Fourier transform of a *Sinc* function is the top hat function, which is unit for absolute wavenumber less than $\pi/d$. The *Sinc* function makes an orthonormal basis on the lattice, where

$$\delta_{nm} = \int_{-\infty}^{\infty} Sinc(x/d - n) Sinc(x/d - m) dx \tag{2}$$

Furthermore, the *Sinc* basis elements are zero at all lattice sites except their own. There are two equivalent but distinct methods for approximating a function; i) the inner product of the function with the *Sinc* basis

$$f_n = \int_{-\infty}^{\infty} F(x) Sinc(x/d - n) dx \qquad f(x) = \sum_{n=-\infty}^{\infty} f_n Sinc(x/d - n) \tag{3}$$

which defines the Eikon; or ii) we can define a sampling of the function with a Dirac-Delta bi-orthogonal with the *Sinc* functions

$$\hat{f}_n = \int_{-\infty}^{\infty} F(x) \frac{\delta(x/d - n)}{\sqrt{d}} dx = F(n)\sqrt{d} \qquad \hat{f}(x) = \sum_n F(n) Sinc(x/d - n) \tag{4}$$

We have found that the integral over all samplings of the Delta-Sinc method equals the Eikon method. Furthermore, we notice that the Eikon of a function is its low pass filtered image at the lattice sites. Decreasing d will monotonically converge the Eikon or the Delta-Sinc sampling to the true function. Furthermore, the Delta-Sinc sampling method adds a complexity to the picture; the origin of the lattice grid can change the sampling. This is not true with the Eikon because it is implicitly a low pass image of the function.

Shannon showed that for a band limited sampled function, all the Eikons are identical and $f_n = \hat{f}_n$ and $f(x) = F(x)$. This was done by showing that the band limited Fourier Transform has a Fourier series equal to the sampling of the function. Therefore, we can represent the function exactly as,

$$F(x) = \sum_{n=-\infty}^{\infty} F(n) \frac{Sin(\pi(x/d - n))}{\pi(x/d - n)} \tag{5}$$

For any *d* such that F has no Fourier wavelengths shorter than *2d*. This is exactly Shannon's result. Therefore, the Eikon is exact and the reconstruction in the *Sinc* basis is perfect for a band-limited functions.

**Inner Product:** The continuous inner product of Eikons is the same as their sampling. The continuous inner product is

$$\langle F|G \rangle = \int_{-\infty}^{\infty} F^*(x) G(x) dx \tag{6}$$

Where Eikons, *F* and *G*, are specified by *f* and *g*,

$$\begin{aligned}
\langle F|G \rangle &= \int_{-\infty}^{\infty} \sum_{n=-\infty}^{\infty} f_n^* Sinc(x/d - n) \sum_{m=-\infty}^{\infty} g_m Sinc(x/d - m) dx \\
&= \sum_{n=-\infty}^{\infty} \sum_{m=-\infty}^{\infty} f_n^* g_m \int_{-\infty}^{\infty} Sinc(x/d - n) Sinc(x/d - m) dx \\
&= \sum_{n=-\infty}^{\infty} \sum_{m=-\infty}^{\infty} f_n^* g_m \delta_{nm} \\
&= \sum_{n=-\infty}^{\infty} f_n^* g_n
\end{aligned} \tag{7}$$

The integral of the *Sinc* functions are always translations of the continuous function, thereby an integer translation of *Sinc* is a Kronecker delta function on the lattice. The expression is exact and the only requirement is the wavelength cutoff presented by Shannon. Next, we discuss the construction of the derivative.

**Universal Lattice Derivative:** We define the lattice derivative via a series of infinitesimal translates of the Eikon. This procedure leads to semi-local derivative matrix that we use to construct the derivative of an Eikon. Translation of a lattice image by an infinitesimal is possible whenever Shannon's criterion is satisfied. In this way the Eikon's derivative can be properly defined as a continuous function. Referring to Equation (5), we can use this formula to effectively translate the Eikon on the lattice. The Eikon is be translated via the operator,

$$T_{nm}(a) = Sinc(n - m + a/d) \tag{8}$$

where $\mathbf{T}(a)$ operating on the Eikon $\mathbf{f}$ give

$$\mathbf{T}(a)f(x) = \sum_m T_{nm}(a) f_m Sinc(x/d - n) = f(x - a) \tag{9}$$

This translation is simply interpolating the Eikon at every point. Notice that this process does not change the origin of the Delta-Sinc sampling. In this way one can shift the Eikon without shifting the lattice itself. We will use the translation matrix to derive the lattice derivative. The derivative by infinitesimal translations is

$$\mathbf{D} = \lim_{\varepsilon \to 0} \frac{\mathbf{T}(\varepsilon) - \mathbf{T}(-\varepsilon)}{2\varepsilon} \tag{10}$$

where the limit can be taken without respect to any particular Eikon. Continuing the process we can show that the universal derivative matrix $\mathbf{D}$ is,

$$D_{nm} = \frac{1}{d} \begin{cases} 0, & n = m, \\ \frac{(-1)^{n-m}}{n - m}, & n \neq m, \end{cases} \tag{11}$$

The expression for $\mathbf{D}$ is semi-local, since an isolated single nonzero lattice element is continuously a Sinc function, which is semi-local but zero on the lattice sites. In an independent work, Fornberg [10] derived this expression from an infinite limit of an finite order difference method. The universal derivative is necessarily related to the translation operator via,

$$\frac{d\mathbf{T}(x)}{dx} = \mathbf{D}\mathbf{T}(x) \tag{12}$$

This relationship has been studied in a context of Shannon wavelets and explored by Cattani [11], where he called the matrix elements of $\mathbf{D}$ connection coefficients. For now, we have put aside the wavelet analysis, but in a future paper we will explore this issue. The *Sinc* function is Cattani's scaling basis. Later we will introduce a periodic space that also satisfies this relationship for the $\mathbf{T}^P$ and $\mathbf{D}^P$ operators. Kosloff in reference [6] also discusses this differential form for the derivative operator. However, Kosloff considered general spaces that are not simple in the inner product, thereby reducing the utility.

**Commutation Relations:** We show that the commutation relationship of the derivative and the position operator is correct on the lattice up to a factor describing the Fourier component on the edge of Shannon's criterion. Whenever this commutation relationship is correct the true continuum derivative can be properly calculated by the lattice derivative. To continue, from Calculus the continuous derivative satisfies the product rule. Specifically on a continuous domain,

$$[\partial/\partial x, x] F(x) = F(x) \tag{13}$$

Let us now consider the lattice derivative: The sampling of the derivative is equal to the derivative matrix acting on the sampled function. We propose to check this correspondence by applying the commutator in Equation (13). We will derive the commutation relation of the derivative and position operator analytically for the Eikons. We represent **D** and **X,** where

$$X_{nm} = d\ n\ \delta_{nm} \tag{14}$$

and **D** is the derivative matrix. Let us compute the commutator,

$$\begin{aligned}[\mathbf{D},\mathbf{X}]_{ij} &= \sum_k D_{ik} X_{kj} - X_{ik} D_{kj} \\ &= \sum_k \frac{(-1)^{i-k}}{i-k} k \delta_{kj} - i \delta_{ik} \frac{(-1)^{k-j}}{k-j} \\ &= \frac{(-1)^{i-j}}{i-j} j - i \frac{(-1)^{i-j}}{i-j} \\ &= \begin{cases} 0 & i = j \\ -(-1)^{i-j} & i \neq j \end{cases} \end{aligned} \tag{15}$$

Which allow us to write the commutator as

$$[\mathbf{D},\mathbf{X}] = \mathbf{I} - \mathbf{A} \tag{16}$$

where

$$A_{ij} = (-1)^{i-j} \tag{17}$$

We refer to **A** as the alternating matrix. It is instructive to investigate the eigenvalues and eigenvectors of the commutator. The ideal commutator should be $n$ degenerate with eigenvalues one. Our commutator constructed from the lattice derivative has $n-1$ eigenvalues of value one and one eigenvalue with value $-(n-1)$. This should be contrasted with the finite difference derivative that do not have this property. We will explore this issue in the next section.

**Alternating Matrix:** The alternating matrix is the only factor in determining all errors to the Eikon. Let us begin with mathematical considerations; all of the eigenvectors of the alternating matrix are zero except one. The non-zero eigenvector of the alternating matrix is $a_n = (-1)^n$. Therefore, the alternating matrix is a projection matrix onto the alternating vector, $\alpha$. The calculation of **Af** is therefore trivially equal to the alternating sum of the terms of the lattice vector times the alternating vector, $\alpha$. In Fast Fourier Transform language, $\alpha$ is the shortest wavelength mode of a space with an even number of elements, which corresponds to a wavelength of *2d*. **A** can be considered an approximation to the amplitude of the Fourier transform at the edge of Shannon's sampling cutoff. However, there is a stability problem. For a symmetric function could be zero under **A**. To fix this problem, we simply calculate the maximum value of the expectation value of **A** under all variations of the origin of the coordinate system. This allows us to define

$$\langle \mathbf{A} \rangle = \underset{a \in [0,d]}{\text{Max}} \left[ \sum_{ijk} f_i T_{ij}(a) A_{jk} T_{ki}(a) f_i \right] \tag{18}$$

We define this quantity to be $\langle \mathbf{A} \rangle$ for the purposes of this paper. All functions with trivial structure above the sampling limit will have an error on their state proportional to $\langle \mathbf{A} \rangle$, which will be demonstrated for several examples.

**Error in Representation:** To demonstrate the utility and accuracy of this novel method, we present the chi-squared error on the representation of a unit Gaussian and the derivative of a unit Gaussian. We sample a unit Gaussian and its derivative onto a variable sized lattice and then compute the error integral, which are given below

$$\chi_0^2(d) = \int_{-\infty}^{\infty} \left| F(x) - \sum_n f_n \text{Sinc}(x/d - n) \right|^2 dx$$

$$= \int_{-\infty}^{\infty} |F(x)|^2 dx - \sum_n |f_n|^2$$

$$\chi_1^2(d) = \int_{-\infty}^{\infty} \left| \frac{dF(x)}{dx} - \sum_n \left( \sum_m D_{nm} f_m \right) \text{Sinc}(x/d - n) \right|^2 dx \tag{19}$$

$$= \int_{-\infty}^{\infty} \left| \frac{dF(x)}{dx} \right|^2 dx - \sum_n \left| \sum_m D_{nm} f_m \right|^2$$

We consider and compare both a finite difference and the lattice derivative. As can be seen in Figure 1, which is a log plot of the square root of the chi-squared error in the representation with respect to decreasing lattice spacing, the interpolation of the unit Gaussian decreases quadratically with decreasing lattice spacing. This is significant and is because The Fourier components of the Gaussian decrease quadratically (since the Fourier transform of a Gaussian is a Gaussian). Also in Figure 1 we show the interpolation of derivative of the function as well as in comparison of the $8^{th}$ order finite difference derivative [12]. Again, our lattice derivative performs significantly better then the finite difference formula. We also see that the error on the derivative is larger by a predictable degree. We have also included $\langle \mathbf{A} \rangle$ to show it tracks above the derivative and its interpolation. The universal derivative acts on all Fourier information encapsulated onto the lattice, and only fails on the Fourier components not encapsulated, which have wavelengths of *2d* or smaller. Again we reiterate that for a band limited function, this expansion is exact.

This new basis defines a unique Hilbert space up to a maximum cutoff wavelength. Derivatives are defined as well as the inner product in a compact analytic form. The continuum of a function is recoverable by Shannon's interpolation formula, Equation (5), in so far that a Kronecker Delta on the lattice becomes a *Sinc* in the continuum. We call this the Universal Lattice Space. In the next section we will consider the extension of this methodology to the periodic domains.

## III. Periodic Domains

**Definition:** Next we will consider Eikons of periodic functions. We will treat these Eikons as functions in all of space. We find that the lattice sum of the Sinc functions is conditional but convergent. We will introduce the summation of the *Sinc* basis, which defines the interpolating basis of the periodic space. All results in the Lattice Universals Section are equally true herein.

**Interpolation of the Periodic Domain:** Extending the *Sinc* basis function to the periodic domain we obtain a simple analytic functions. These functions are defined on a *2n+1* space of lattice points. This is equivalent to the method outlined for the interpolation in Fourier space, which was considered as early as 1800's for harmonic approximations to planetary motion, Heideman [13] gives the history of the harmonic approximation. To continue, it is necessary to divide the interpolation procedure between even and odd spaces. The even space is ill defined because it contains the *2d* alternating wavelength mode, which is not allowed by the sampling theorem. The odd space is orthogonal and is defined by the basis, $B^P$, written below,

$$B^P(\vartheta) = \sum_{l=-\infty}^{\infty} Sinc(\vartheta + lN) = \begin{cases} 1, & \vartheta \bmod N = 0, \\ \frac{1}{2N}\left(Tan(\frac{\pi\vartheta}{2N}) + Cot(\frac{\pi\vartheta}{2N})\right) Sin(\pi\vartheta), & else \end{cases} \quad (20)$$

and where $B^p$ is the kernel of the periodic space translation operator,

$$T^P_{nm}(a) = B^P(n - m + a/d) \quad (21)$$

where $T^P$ is the corresponding translation matrix defined in the same way as Equation (9). The translation matrix provides a means to interpolate the periodic domain. We will define the periodic derivative in the next section.

**Periodic Derivative:** The periodic lattice derivative is shown to be an analytic trigonometric function from the derivative limit of small translations, which proceeding from Equation (10). The limit has been taken to form an analytic expression:

$$D^P_{nm} = \frac{1}{d}\begin{cases} 0, & n = m, \\ (-1)^{n-m}\frac{\pi}{N} Csc(\frac{\pi(n-m)}{N}), & else \end{cases} \quad (22)$$

Also, the same completeness relationship, Equation (12), applies to the periodic derivative and translation operator. This derivative has been found in the literature independently in Fornberg [10], which was derived from a derivative acting on the harmonic Fourier interpolation formula.

**Comparison with Fast Fourier Transforms:** We find that the periodic lattice derivative and the derivative defined by the Fast Fourier transform are equivalent. For a periodic lattice from $[-N,N]$ elements, let the Discrete Fourier Space basis be defined by $e_k$ for with wavelength $\lambda_k$. The commonly defined lattice derivative using Fast Fourier Transform is then

$$\mathbf{D}^{FFT} = \sum_{k=1}^{N} \frac{2\pi}{\lambda_k} \left( \mathbf{e}_k \otimes \mathbf{e}_k^* \right) \tag{23}$$

where

$$\lambda_k = N/k \qquad e_k(n) = Exp[i2\pi n/\lambda_k]/\sqrt{N} \tag{24}$$

which is equivalent to the works of Marston [4] and Kosloff [5, 6]. Using equations (20), (23) and (24), it can be shown the equivalence of the derivative operators. Therefore, the periodic phase space expression for the lattice derivative is equivalent to Equation (20) and the eigen-system of $\mathbf{D}^P$ is identical to the Fast Fourier Transform eigen-system.

**Error in Representation:** We have compared directly the unit Gaussian for the periodic space as well. All compact functions have been found to behave similarly in both bounded and periodic spaces. We have applied the same definitions to the problem with an odd number of lattice sites. In Figure 2 we report the identical scaling to that found in the bounded case in the previous section. The differences in the two derivatives are not consequential in this analysis because the Gaussian is compact.

## IV. Application to Quantum Mechanics

**Hermitian Operator:** The $\mathbf{D}$ matrix is proportional to the representation of the momentum operator in Quantum Mechanics, therefore we define

$$\mathbf{P} = -i\hbar \mathbf{D} \tag{25}$$

The Hermitian operator $\mathbf{P}$ is the generator of translations on the lattice and is a common construction in Quantum Mechanics.

**Canonical Commutation Relations:** Using Equation (16), we evaluate the canonical commutations relationships used in Quantum Mechanics for bounded domains,

$$[\mathbf{X}, \mathbf{P}] = i\hbar (\mathbf{I} - \mathbf{A}) \tag{26}$$

This shows that band limited functions that nominally satisfy calculus will also satisfy relationships of Quantum Mechanics.

**Schrödinger Wave Equation:** From Equation (25) we can construct the Schrodinger Wave equation for quantum mechanical problems in our Eikon method using a Hamiltonian approach,

$$\mathbf{H} = \left( \frac{\mathbf{P}^2}{2m} + \mathbf{V} \right) \tag{27}$$

where one finds the eigenvalues of the Hamiltonian, $\mathbf{H}$, to get the energy spectra of the quantum system.

## V. Examples and Results

**One- Dimensional Morse potential:** As our first example, we consider the one-dimensional Morse potential from reference [14]. We solved this problem using our sampling method with 129 lattice elements. In comparison, Marston et. al. [4] have solved this problem using Fourier techniques and obtained errors in the order of $10^{-6}$. Our results are listed in Table 1, where we used physical constants consistent to five significant digits with Marston. The improvement over the results in reference [4] we attribute to the improved quality of the eigenstate calculations available. The same technique with a periodic derivative operator gives similar results.

**Quartic Anharmonic Oscillator:** As another one-dimensional example of our new techniques, we solve the quartic anharmonic oscillator. In reference [15] Tymczak et. al. has solved the quartic anharmonic oscillator to very high precision which we will use for comparison. The Hamiltonian for the quartic anharmonic oscillator is,

$$\mathbf{P}^2 - Z^2\mathbf{X}^2 + \mathbf{X}^4 \tag{30}$$

We solve this problem with our sampling technique where we set $d = 0.15$ and $\mathbf{N} = 90$. In Table 2 we present our calculations of the anharmonic oscillator in comparison to the results of reference [15]. As can be seen in Table 2, we obtain extremely accurate results for our numerical sampling method. Also, we have calculated the scaling relationship for the ground state energy of the $Z=0$ anharmonic oscillator. In Figure 3 we present the parabolic scaling towards the correct answer. This again shows that all Fourier Components are treated perfectly up to the error on the edge of the Fourier space. We find quadratic convergence like the interpolation integrals preformed in the bounded domain.

**Coulomb Potential and the Hydrogen Atom:** Given the success of the proceeding examples, we will now demonstrate the capabilities of our Eikon method on a computationally challenging and technologically important problem, the Hydrogen Atom. We begin by finding the Eikon of the coulomb potential. The eikon of the coulomb potential, V, is effectively the low pass image at a scale $d$,

$$V(\vec{r}) = \sum_{\vec{n}} \frac{Ir_{\vec{n}}}{d} \frac{\sin(\pi(x/d - n_x))}{\pi(x/d - n_x)} \frac{\sin(\pi(y/d - n_y))}{\pi(y/d - n_y)} \frac{\sin(\pi(z/d - n_z))}{\pi(z/d - n_z)} \tag{31}$$

Where Ir is the "inverse of radius" lookup table that is calculated beforehand. The calculation of Ir can be conducted on a finite domain by transforming the *Sinc*'s into Fourier space, which means we represent the *Sinc's* in terms of the Fourier integral.

$$\frac{Ir_{\vec{n}}}{d} = \frac{1}{d^3}\int_{-\infty}^{\infty} dx\,dy\,dz \frac{1}{\sqrt{x^2+y^2+z^2}} \frac{\sin(\pi(x/d - n_x))}{\pi(x/d - n_x)} \frac{\sin(\pi(y/d - n_y))}{\pi(y/d - n_y)} \frac{\sin(\pi(z/d - n_z))}{\pi(z/d - n_z)}$$

$$= \frac{1}{d^3}\int_{-\infty}^{\infty} d^3r \frac{1}{|\vec{r}|}\int_{-\pi}^{\pi}\frac{d\beta^3}{(2\pi)^3} Exp[-i(\vec{\beta}\cdot\vec{r})/d + i(\vec{\beta}\cdot\vec{n})] \qquad (32)$$

$$= \frac{1}{d}\int_{-\pi}^{\pi}\frac{d\beta^3}{(2\pi)^3}\frac{4\pi}{\beta^2} e^{i\vec{\beta}\cdot\vec{n}}$$

As can be seen Ir is independent of *d*. The coulomb potential is scale free; therefore the only scale in the entire problem is the sampling distance *d*. In Table 3 we give the first few values of the Coulomb potential used in the calculation. The next step is to convert this one point *sinc* integral into a two-point off-diagonal *sinc* integral. We noticed that any pair of *sincs* has a minimum Fourier wavelength of *d*, which corresponds to twice the sampling rate of *d*. We can therefore complete the two point coulomb-sinc integral by oversampling V with a sampling distance of *d/2*. This allows us to build the Coulomb matrix.

All results reported have been calculated in a ten atomic unit box. All calculations are in terms of atomic units; therefore the final energy for Hydrogen is minus one half Hartree, the kinetic term is plus one half Hartree and the Potential term is minus one Hartree. We can conduct these calculations quickly using LAPACK eigenvalue package on a symmetric real matrix [16], where we select the particular solutions to be found and it takes less than ten seconds to calculate the $15^3$ lattice site ground state.

In Table 4 we report the energies of the ground state with decreasing lattice spacing. We calculate 1% errors on the energy of the ground state in a $21^3$-lattice site calculation. We plot a cross section of the wave function through the origin in Figure 5. To demonstrate the sensitivity of the ground state energy to the choice of the grid we translated the Coulomb potential, which is shown in Table 5. As shown in Table 5 we find 0.02% variations in the energy under varied translations at $15^3$ lattice sites.

For instructive purposes, we compare these results to a wavelet techniques demonstrated by Tymczak et. al. in reference [17] (for a review of wavelet techniques in electronic structure see [18-24]). Tymczak reports 2% errors on an analysis of the same Hydrogen system. Tymczak demonstrated that with a sparsely filled $128^3$-lattice grid results were comparable to the results on this methods $15^3$-lattice site calculation. However, this method's computed kinetic and potential energies are much more accurate than Tymczak's wavelet analysis. In a following publication we will extend this methodology to include multiply atoms at the Hatree-Fock level of theory and show that using a mixed basis set we can obtain very accurate computation of ground state energies and properties [25].

## VI. Conclusion

We have introduced an approach based on Shannon's Sampling theorem to solve Quantum Mechanical systems. We have shown that by extending the logic of Shannon's interpolation theorem we can define a Universal Lattice Basis that has superior interpolating properties compared to tradition methods. We showed that this basis is orthonormal, semi-local, has a Euclidean norm and a simple analytic expression for the derivatives. Additionally, we defined a bounded domain for which band-limited functions, such as Gaussians, show quadratic convergence in the representation error in respect to the sampling frequency. We were also able to extend this theory to the periodic domain and illustrate the simple analytic forms of the periodic semi-local basis and derivatives. Additionally, this periodic basis is equivalent to the space defined by the Fast Fourier Transform. We have shown that this novel basis has great utility in solving quantum mechanical problems for which the wave functions are known to be naturally band-limited. The first simple one-dimensional example that we demonstrated was the Morse potential. We were able to compute the ground state wave functions and energies of the Morse potential to a precision of 8 digits or greater using 129 basis functions. We then used this method to solve for the an-harmonic oscillator and again obtain a precision of grater than 10 digits. Next we extended this method to study a three-dimensional singular potential problem, the Hydrogen atom. Here, because of the singular nature of the potential, we where able to obtain a 3-4 digit precision in the ground state energy with a grid of $21^3$ basis functions, which is substantially more precise that grid based methods can obtain at this grid density. As a side note, we are in the process of developing this method for atomic and molecular systems, which will be the subject of a later publication [25].

## VII. Acknowledgements

The authors would like to acknowledge the support given by Welch Foundation (Grant J-1675), the ARO (Grant W911Nf-13-1-0162), the Texas Southern University High Performance Computing Center (http:/hpcc.tsu.edu/; Grant PHY-1126251) and NSF-CREST CRCN project (Grant No. 1137732). Dr. Jerke would also like to thank University of Houston for hosting him as a Visiting Assistant Professor while he developed the kernel of this idea.

## VIII. List of Figures

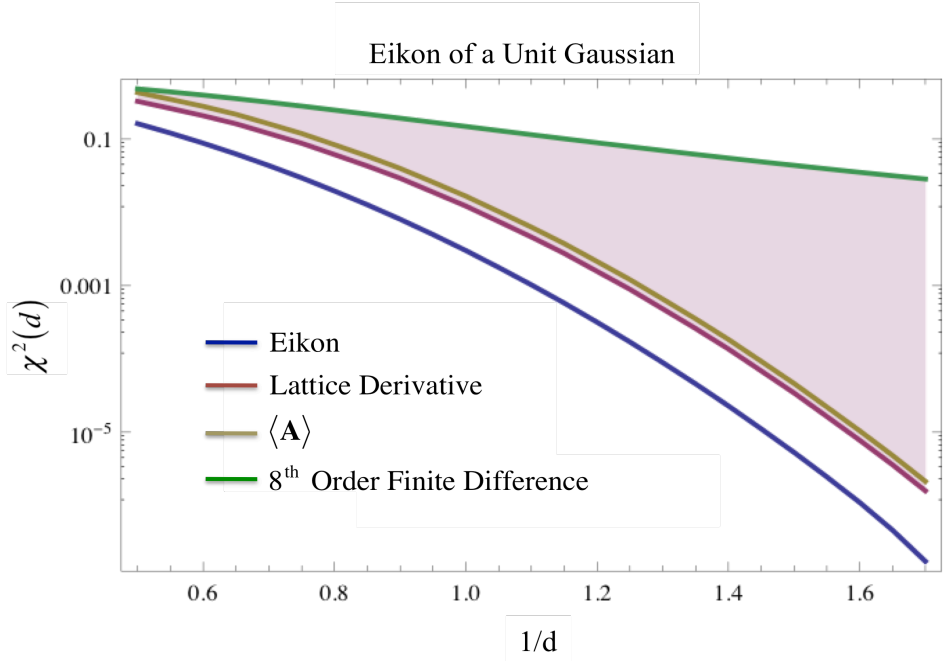

**Figure 1:** shows the $\chi^2$ of the fits defined in Equation (16). The errors reduce quadratically because the function is a Gaussian, which is band limited. We also show the expectation value of the commutator, which is an approximate upper bound to the error.

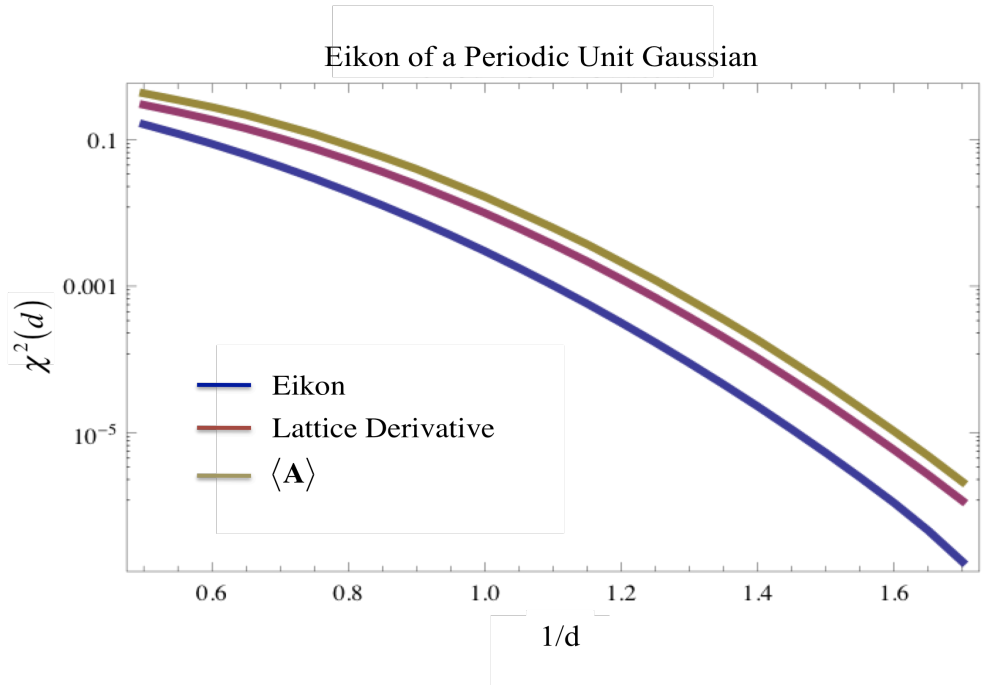

**Figure 2:** shows the $\chi^2$ of the fits defined in Equation (16) for the periodic case. For compact functions like a unit Gaussian, the periodic and bounded systems have identical behavior.

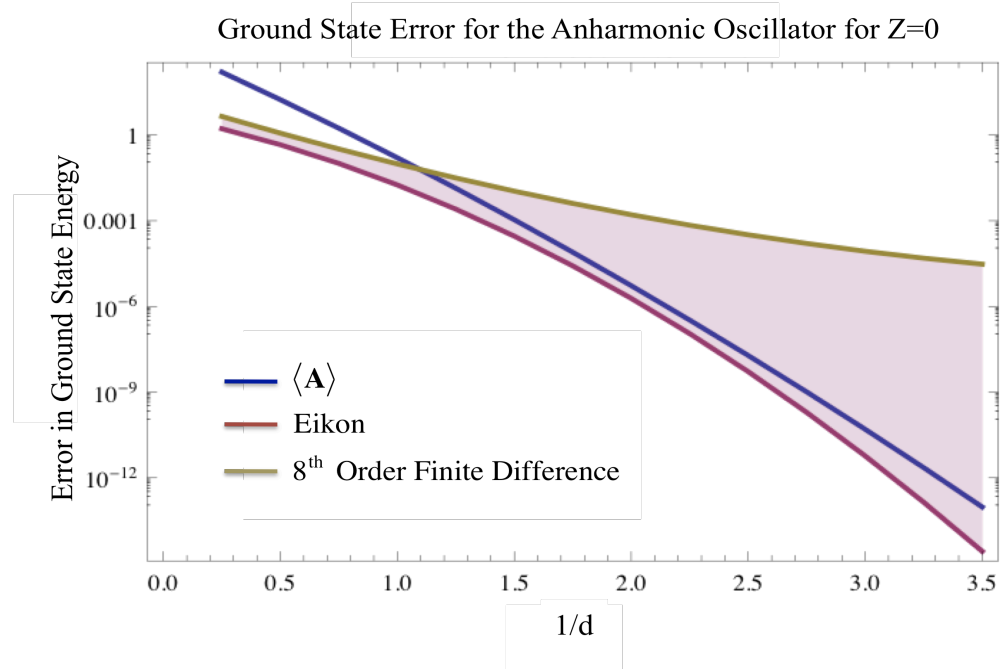

**Figure 3:** Shows the comparison of the ground state energies errors of the quartic anharmonic oscillator, Equation (30). A finite difference matrix was used as a comparison to the technique introduced in this paper. The expectation value of **A** can be used to estimate a bound in the error of the ground state energy

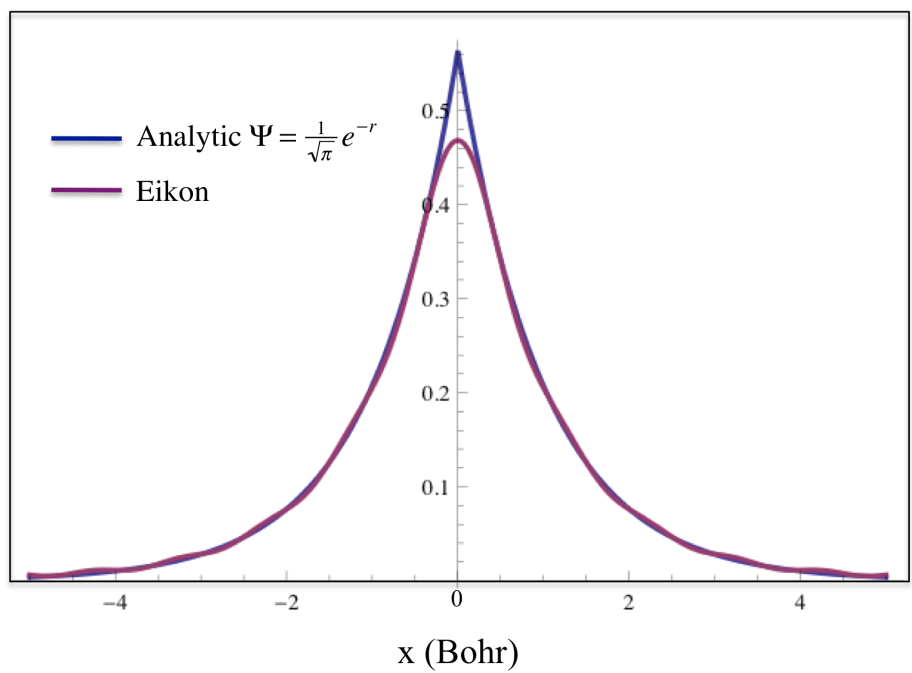

**Figure 4:** A cross section of the Hydrogen wave function amplitude at $21^3$ lattice sites using a 10 au box. We interpolated the result along the line through the origin. The analytic function is also plotted as a comparison.

## IX. List of Tables

**Table 1:** The energy spectra of the Morse Potential calculated with 129 lattice sites. All units are in atomic units. Notice that we only know the physical parameters of the Morse potential to five significant digits.

| State Number | Fourier Method | Eigen Method | Analytic |
|---|---|---|---|
| 0 | 0.009868818 | 0.009868818 | 0.009868818 |
| 1 | 0.028744212 | 0.028744212 | 0.028744212 |
| 2 | 0.046469952 | 0.046469952 | 0.046469952 |
| 3 | 0.063046036 | 0.063046036 | 0.063046036 |
| 4 | 0.078472465 | 0.078472465 | 0.078472465 |
| 5 | 0.092749239 | 0.092749239 | 0.092749239 |
| 6 | 0.105876358 | 0.105876358 | 0.105876358 |
| 7 | 0.117853822 | 0.117853822 | 0.117853822 |
| 8 | 0.128681631 | 0.128681631 | 0.128681631 |
| 9 | 0.138359785 | 0.138359785 | 0.138359785 |
| 10 | 0.146888283 | 0.146888283 | 0.146888283 |
| 11 | 0.154267127 | 0.154267127 | 0.154267127 |
| 12 | 0.160496315 | 0.160496315 | 0.160496315 |
| 13 | 0.165575848 | 0.165575848 | 0.165575848 |
| 14 | 0.169505726 | 0.169505726 | 0.169505726 |
| 15 | 0.172285949 | 0.172285949 | 0.172285949 |
| 16 | 0.173916549 | 0.173916541 | 0.173916517 |

**Table 2**: The first and second Eigenstates of the quartic anharmonic oscillator as a function of Z. We compare our results to Tymczak et. al. who solved this problem to an accuracy of 30 digits.

| | Numerical | | Analytical | |
|---|---|---|---|---|
| $Z^2$ | 1$^{st}$ Eigenvalue | 2$^{nd}$ Eigenvalue | 1$^{st}$ Eigenvalue | 2$^{nd}$ Eigenvalue |
| 0 | 1.060362090 | 3.799673030 | 1.060362090 | 3.799673030 |
| 1 | 0.657653005 | 2.834536202 | 0.657653005 | 2.834536202 |
| 5 | -3.410142761 | -3.250675362 | -3.250675362 | -3.410142761 |
| 15 | -20.633576703 | -20.633546884 | -20.633576703 | -20.633546884 |
| 20 | -50.841387284 | -50.841387284 | -50.841387284 | -50.841387284 |
| 25 | -149.219456142 | -149.219456142 | -149.219456142 | -149.219456142 |

**Table 3:** The low pass image of the Coulomb potential evaluated numerically. The results are independent of the lattice spacing; instead they scale inversely with the lattice spacing. This is because the Coulomb potential is scale free. We notice that the central Eikon value is approximately the diffraction limited angular scaling factor times two. We calculated the zero state exactly.

| Element | Value |
|---------|-------|
| (0,0,0) | 2.442749607806 |
| (1,0,0) | 1.05169 |
| (1,1,0) | 0.72669 |
| (1,1,1) | 0.58508 |
| (2,0,0) | 0.47399 |

**Table 4:** The ground state energies of Hydrogen due to the Coulomb potential. The number of lattice sites and the lattice spacing in terms of atomic units are reported; all calculations have a lattice size of 10 au. The total, kinetic, and potential energies are reported.

| N | d (au) | E (Hartree) | T (Hartree) | V (Hartree) |
|---|--------|-------------|-------------|-------------|
| $5^3$ | 2.500 | -0.385 | 0.228 | -0.613 |
| $9^3$ | 1.250 | -0.466 | 0.396 | -0.862 |
| $11^3$ | 1.000 | -0.479 | 0.434 | -0.914 |
| $15^3$ | 0.714 | -0.49065 | 0.472 | -0.963 |
| $21^3$ | 0.500 | -0.49595 | 0.492 | -0.988 |

**Table 5:** The variation in the energy of the Hydrogen under translation of the Coulomb potential. The lattice spacing is 0.714 atomic units and the number of lattice sites is 15 on a side. We parameterize the move of the potential in terms of the distance from the origin in atomic units.

| Shift ( au ) | E (Hartree) |
|--------------|-------------|
| 0.071 | -0.49065 |
| 0.143 | -0.49057 |
| 0.214 | -0.49051 |
| 0.286 | -0.49045 |
| 0.357 | -0.49042 |